\documentclass[aip,
reprint,
groupedaddress,
]{revtex4-1}
\usepackage[svgnames]{xcolor}
\usepackage{amsmath}
\usepackage{bm}
\usepackage{braket}
\usepackage{listings}
\usepackage{amssymb}
\usepackage{graphicx}
\usepackage{comment}
\usepackage{multirow}
\graphicspath{{./img/}}

\DeclareMathOperator{\tr}{tr}

\renewcommand{\Re}{\operatorname{Re}}

\begin{document}
\title{Optical response of laser-driven charge-transfer complex described by Holstein--Hubbard model coupled to heat baths: Hierarchical equations of motion approach}
\author{Kiyoto Nakamura}
\email{nakamura.kiyoto.n20@kyoto-u.jp}
\author{Yoshitaka Tanimura}
\email{tanimura.yoshitaka.5w@kyoto-u.jp}
\affiliation{Department of Chemistry, Graduate School of Science, Kyoto University, Sakyoku, Kyoto 606-8502, Japan}
\date{\today}	

\begin{abstract}
We investigate the optical response of a charge-transfer complex in a condensed phase driven by an external laser field. Our model includes an instantaneous short-range Coulomb interaction and a local optical vibrational mode described by the Holstein--Hubbard (HH) model. Although characterization of the HH model for a bulk system has typically been conducted   using a complex phase diagram, this approach is not sufficient for  investigations of  dynamical behavior at finite temperature, in particular for  studies of nonlinear optical properties, where the time irreversibility of the dynamics that arises from the environment becomes significant. We therefore include heat baths with  infinite heat capacity in the model to introduce thermal effects characterized by fluctuation and dissipation to the system dynamics. By reducing the number of degrees of freedom of the heat baths, we derive numerically ``exact'' hierarchical equations of motion (HEOM) for the reduced density matrix of the HH system. As demonstrations, we   calculate the optical response of the system in  two- and four-site cases under external electric fields.The results indicate that the effective strength of the system--bath coupling becomes large as the number of sites increases. Excitation of electrons promotes the conductivity when the Coulomb repulsion is equivalent to or dominates the electron--phonon coupling, whereas excitation of optical vibrations always suppresses the conductivity.
\end{abstract}
\maketitle


\section{INTRODUCTION}
In various organic conductors such as  bis(ethylenedithio)tetrathiafulvalene (BEDT-TTF)\cite{BEDT-TTF} and the fullerides (salts of C$_{60}$ anions),\cite{C60} the interplay of on-site electron--electron and on-site phonon--electron interactions plays a significant role. Owing to the competition between these two interactions, such materials exhibit a variety of distinct states in the antiferromagnetic (AF), charge-density-wave (CDW), spin-density-wave (SDW), and superconducting (SC) phases. For a description of the complex phase diagram of  correlated electron systems, the Holstein--Hubbard (HH) model has been introduced.\cite{holstein59-1, holstein59-2, beni74hh, aubry95exact, takada03hh, sykora08hh, miyao18hh, berciu07hh, yamada14hh}
This model consists of electrons on a discrete lattice, with on-site Coulomb repulsion between electrons and a local coupling of electrons to longitudinal optical (LO) phonons. It is characterized by the transfer integral $t_h$, the on-site Coulomb repulsion $U$, the  LO phonon frequency $\Omega$, and the electron--phonon coupling strength $g$.

For the half-filled HH model, when the Coulomb repulsion $U$ is much larger than the electron--phonon coupling $g$, two electrons with  opposite spin directions cannot occupy the same site, and electrons with up spin and those with down spin are aligned in an alternating sequence in the Mott AF phase, whereas when $g$ is much larger than $U$, two electrons with opposite spin directions occupy the same site with the aid of an LO phonon, and bipolarons are formed in the Peierls CDW phase. Then, between the AF and CDW phases, the existence of a SC state is predicted, which is responsible for the high-temperature SC state of alkali fullerides.\cite{hebard91c60, rosseinsky91c60, palstra95c60} The role of phonons in the SC state of cuprates has also been investigated along the same lines,\cite{lanzara01cuprate, huang03cuprate, johnston12cuprate, ishihara04cuprate, mishchenko08cupurate, de09cupurate} and the HH model, despite  its simplicity, does seem  to represent a versatile approach for the investigation of these materials. 

Many numerical studies, based, for example, on  exact diagonalization, \cite{takada96hh, chatterjee10hh} the density matrix renormalization group (DMRG), \cite{Tezuka07PRB} and the quantum Monte Carlo (QMC) method, \cite{von95hh, macridin04hh, clay05hh, nowadnick12hh, johnston13hh, Ohgoe17PRL, weber18hh} have been conducted to reveal the thermodynamic properties of the HH model at zero and finite temperatures.  Most of these investigations have focused on bulk systems,  typically through the use of a complex phase diagram, and, with a few exceptions (see, e.g., Ref.~\onlinecite{tsuji09nonequilibrium}), there has been little exploration of the  dynamical behavior of the HH model. The recent discovery of SC-like properties initiated by optical pulses and observed at  temperatures higher than the SC critical temperature $T_c$ provides  considerable motivation for  dynamical investigations of the HH model,\cite{Mitrano16Nat} as do the current efforts to achieve room-temperature superconductivity with the aid of optical pulses, in the context of which both theoretical  \cite{mazza17PRB, sentef16PRB, kennes17NatP, Kaneko19PRL, ido17num, kim16num, werner15num, golez12num} and experimental\cite{cantaluppi18exp, nicoletti17exp, mankowsky17exp} investigations have already been conducted.

 To investigate dynamical behavior of an HH system, not only the LO phonon modes but also environmental degrees of freedom have to be taken into account, because the latter control the system temperature and the time scales of relaxation processes of excited states through  fluctuations and dissipation  related through the fluctuation--dissipation theorem. However, investigations of dynamics, in particular investigations of nonlinear optical properties at finite temperature, have not been conducted thoroughly, even for a small cluster system. 

In this paper, we develop an open quantum dynamics theory for the HH model, in which the HH system is coupled to baths consisting of surrounding atoms or molecules. Here, by an open quantum dynamics theory we mean a theory pertaining to a system that is further coupled to bath systems that are  modeled by  infinite numbers of harmonic oscillators.\cite{weiss12book}
Such system--bath models have been employed for investigations of electron transfer\cite{Onuchic} and exciton transfer processes\cite{Ishizaki2009, Hein_2012}  of a molecular system, in particular for calculations of the nonlinear optical spectrum.
However, the effects of on-site Coulomb repulsion between electrons, which plays an essential role in charge transfer complexes, has not been  investigated in any great depth.

We derive hierarchical equations of motion (HEOM) for the reduced density matrix of the HH system that can treat non-Markovian and nonperturbative system--bath interactions at finite temperature in a numerically ``exact'' manner. \cite{heomHigh, TanimuraJPSJ06, tanimuraJCP20} The HEOM approach has been applied to  the Holstein\cite{LipenHolns2015,ZhaoPolaron,ReichmanHolns2019} and Peierls--Holstein\cite{Mauro2021} problems.  Photosynthetic systems have also been investigated using this  approach, and it has been shown that the efficiency of the exciton transfer process exhibits a maximum in the nonperturbative region of  system--bath coupling,\cite{Ishizaki2009} owing to the quantum entanglement between the system and bath (``bathentanglement'').\cite{tanimuraJCP20} Using the HEOM approach, we can explore the possibility of finding such a mechanism in this HH system. We illustrate our formalism by calculating the optical conductivity of the one-dimensional HH model at finite temperature.  In addition, by applying time-dependent external fields, we calculate the optical conductivity under highly nonequilibrium conditions. Because the heat bath facilitates  thermal dissipation and excitation in a thermodynamically consistent manner, we can obtain reliable results under such conditions, even when the molecular system described by the HH model is very small.

The remainder of the paper is organized as follows. In Sec.~\ref{sec:model}, we introduce the system--bath Hamiltonian for the HH model. In Sec.~\ref{sec:method}, we present the HEOM for the HH model. Section~\ref{sec:results} is devoted to a presentation of the numerical results for two- and four-site HH-plus-bath models. The optical conductivities of the equilibrium states and the steady states under external driving fields are investigated. We conclude the paper in Sec.~\ref{sec:conclude}.

\section{MODEL} \label{sec:model}
We consider a one-dimensional HH model described by
\begin{align}
  \hat{H}_{S} (t)=&  -\sum_{i, \sigma} (t^i_h e^{iA(t)}
  \hat{c}_{i, \sigma}^{\dagger} \hat{c}_{i+1, \sigma} + \mathrm{H.c.})
  + \sum_{i} U^i \hat{n}_{i, \uparrow} \hat{n}_{i, \downarrow} \nonumber \\[3pt]
  & + \sum_{i}g^i(\hat{n}_{i, \uparrow} + \hat{n}_{i, \downarrow})(\hat{b}_{i}^{\dagger} + \hat{b}_{i})
  +  \sum_{i} \Omega^i \hat{b}_{i}^{\dagger} \hat{b}_{i} \nonumber \\[3pt]
 & - E_{\mu}(t) \sum_{i}(\hat{b}_{i}^{\dagger} + \hat{b}_{i}),
\label{eq:H_S}
\end{align}
where $c_{i, \sigma}$ and $c_{i, \sigma}^{\dagger}$ are the annihilation and creation operators of an electron at  site $i$ with  spin state $\sigma \in \{\uparrow, \downarrow\}$, $b_{i}$ and $b_{i}^{\dagger}$ are the annihilation and creation operators of an LO phonon at  site $i$, and  $t^i_h, U^i, g^i$, and $\Omega^i$ are the transfer integral, on-site Coulomb repulsion, electron--phonon coupling strength, and LO phonon frequency, respectively, at site $i$. The functions $A(t)$ and $E_{\mu}(t)$ are the time-dependent external perturbations for electrons and LO phonons, described by the vector potential and electric field;  the static case is given by $A(t)=0$ and $E_{\mu}(t)=0$.  Because we treat each site explicitly, these parameter values can be site-dependent.  For example, using a machine learning approach, we can set a system-specific parameter value targeting  a specific complex molecular system.\cite{Ueno2021}
The probability distribution of the electron is expressed as $\hat{n}_{i, \sigma} = \hat{c}_{i, \sigma}^{\dagger}\hat{c}_{i, \sigma}$.
Here and in the rest of this paper, we set the reduced Planck constant and the elementary charge equal to $1$, i.e., $\hbar = e = 1$.

In this study, we consider a case in which the LO modes are coupled to intermolecular vibrations or other molecular degrees of freedom, and we describe these degrees of freedom using a harmonic heat bath.
Because we can study only a small system with a finite number of phonon modes associated with the system site, inclusion of the heat bath is important to maintain the stability of the equations.\cite{ReichmanHolns2019}
Thus, we consider the situation in which each LO oscillator interacts with a heat bath that gives rise to dissipation and fluctuation in the LO modes.
The total Hamiltonian is then expressed as
\begin{align}
\hat{H}_\mathrm{tot}(t) = \hat{H}_{S}(t) + \hat{H}_{I} + \hat{H}_{B}.
\end{align}
The bath Hamiltonian $\hat{H}_{B}$ and the system--bath interaction $\hat{H}_{I}$ are expressed as
\begin{align}
\hat{H}_{B} &=  \sum_{i, \alpha}
\left(\frac{\hat{p}_{i, \alpha}^{2}}{2 m_{i, \alpha}} + \frac{1}{2} m_{i, \alpha} \omega_{i, \alpha}^{2} \hat{x}_{i, \alpha}^{2}\right),
\label{H_B}
\\[3pt]
\hat{H}_{I} &=  -\sum_{i} \hat{V}_{i} \sum_{\alpha} k_{i, \alpha} \hat{x}_{i, \alpha}
+ \sum_{i} \hat{V}_{i}^{2} \sum_{\alpha} \frac{k_{i, \alpha}^{2}}{2 m_{i, \alpha} \omega_{i, \alpha}^{2}},
\label{H_I}
\end{align}
where $\hat{p}_{i, \alpha}, \hat{x}_{i, \alpha}, m_{i, \alpha}$, and $\omega_{i, \alpha}$  are the momentum, position, mass, and frequency of the $\alpha$th oscillator of the $i$th bath, respectively.
The system operator that represents the effects of the $i$th local heat bath is given by $\hat{V}_{i}$, and $k_{i, \alpha}$ is the coupling constant for the interaction between the system and the $\alpha$th oscillator of the $i$th bath.
The second term in $\hat{H}_{I}$ is a counterterm that compensates for the renormalization of the potential energy caused by the first term in $\hat{H}_{I}.$\cite{weiss12book}
Depending on the situation, we may choose a local or nonlocal system--bath interaction by the way in which we set $\hat{V}_{i}$: for high-frequency intramolecular vibrational modes, we consider the case in which each LO oscillator is coupled to its own bath, which is expressed as $\hat{V}_{i} = \hat{b}_{i}^{\dagger} + \hat{b}_{i}$,  whereas for low-frequency intermolecular vibrational modes or phonon modes, we consider the case in which multiple LO oscillators are coupled to a single global bath, described as $\hat{V}_i \to \hat{V}_{0} = \hat{b}^{\dagger}_{0} + \hat{b}_{0}$, with  $i$ being ignored in the summations in Eqs.~\eqref{H_B} and \eqref{H_I}.
If necessary, we can further treat a nonlocal heat bath described by $\hat{V}_{ij}$ between the $i$th and $j$th sites, which often plays an important role in determining the efficiency of an energy transfer process.\cite{Mauro2021,Ueno2021}

The $i$th heat bath can be characterized by the spectral distribution function (SDF), defined by
\begin{align}
  J_{i}(\omega) = \sum_{\alpha}^{N_i} \frac{k^{2}_{i, \alpha}}{2 m_{i, \alpha} \omega_{i, \alpha}} \delta(\omega - \omega_{i, \alpha}).
\end{align}
For the heat bath to be an unlimited heat source possessing an infinite heat capacity, the number of  heat-bath oscillators $N_{i}$ is effectively made infinite by replacing $J_i (\omega)$ with a continuous distribution. The counterterm of $\hat{H}_{I}$ can then be rewritten in terms of the reorganization energy  $\lambda_{i}= \int_{0}^{\infty} d\omega \,J_{i}(\omega)/\omega$ as $\sum_{i} \lambda_{i} \hat{V}_{i}^{2}$.

Although we can reduce the number of  phonon degrees of freedom by diagonalizing the LO phonon plus bath degrees of freedom, which leads to the Hubbard model with Brownian-type heat bath, as described in Refs.~\onlinecite{Onuchic,Ueno2021,TaniMukaJPSJ90,TanakaHEOM1,TanakaHEOM2,TJCP2012,Mauro2021}, here we treat the LO phonon modes expressed in the eigenstate representation explicitly to reduce the computational cost. 
In this way, we can investigate the case in which the LO phonon modes are directly excited by a laser field (see Sec.~\ref{subsec:noneq}).
Moreover, if necessary, we can include an anharmonicity of the phonon modes by setting each excitation energy of the phonon state separately without incurring any additional cost.

\section{HEOM FOR THE HH-PLUS-BATH SYSTEM} \label{sec:method}
Our open quantum dynamics theory is constructed on the basis of the reduced density operator, defined as $\hat{\rho}(t) = \tr_{B} \{e^{-i \hat{H}_\mathrm{tot}(t)} \hat{\rho}_\mathrm{tot}(0) e^{i \hat{H}_\mathrm{tot}(t)}\}$, where $\tr_{B}\{~\}$ represents the partial trace over the bath degrees of freedom.

If we adopt the Drude--Lorentz SDF expressed as
\begin{align}
 J_{i}(\omega) = \frac{\eta_{i} \omega}{\pi} \frac{1}{1 + (\omega/ \gamma_{i})^2},
 \label{eq:sdf}
\end{align}
then the time evolution of the reduced density operator is  described by the HEOM\cite{tanimuraJCP20} for the HH-plus-bath model as follows (see Appendix~\ref{sec:appHEOM}):
\begin{widetext}
\begin{align}
  \label{eq:HEOM}
  \frac{\partial}{\partial t} \hat{\rho}_{\vec{n}_{1}, \ldots, \vec{n}_{N}}(t) ={}& 
  - \left(i\hat{H}_{S}^{\times} (t)
  + \sum_{i=1}^{N}\sum_{k=0}^{K} n_{i, k} \nu_{i, k} \right )\hat{\rho}_{\vec{n}_{1}, \ldots, \vec{n}_{N}}(t)\nonumber\\[3pt]
   &-\sum_{i=1}^{N}\left( i\lambda_{i} (\hat{V}_{i}^{2})^{\times} \hat{\rho}_{\vec{n}_{1}, \ldots, \vec{n}_{N}}(t)  
   + i \hat{V}_{i}^{\times} \sum_{k=0}^{K}
  \hat{\rho}_{\vec{n}_{1}, \ldots, \vec{n}_{i}+\vec{e}_{k}, \ldots, \vec{n}_{N}}(t)\right.  \nonumber\\[3pt]
   &\hspace{32pt}+ i  \hat{V}_{i}^{\times} \sum_{k=0}^{K}  n_{i, k} c_{i, k}
  \hat{\rho}_{\vec{n}_{1}, \ldots, \vec{n}_{i}-\vec{e}_{k}, \ldots, \vec{n}_{N}}(t) 
   \left. + \hat{V}_{i}^{\circ} \frac{\eta_{i} \nu_{i, 0}^{2}}{2}
  \hat{\rho}_{\vec{n}_{1}, \ldots, \vec{n}_{i}-\vec{e}_{0}, \ldots, \vec{n}_{N}}(t) \right).
\end{align}
\end{widetext}
Here, the vector $\vec{n}_{i} = (n_{i, 0}, n_{i, 1}, \ldots, n_{i, K})$ consists of nonnegative integers, and $\hat{\rho}_{\vec{0}, \ldots, \vec{0}}(t)$ corresponds to the reduced density operator.
The density operators whose vectors $\{\vec{n}_{i}\}$ involve positive integers are referred to as the auxiliary density operators (ADOs).
The sets $\{c_{i, k}\}$ and $\{\nu_{i, k}\}$ are evaluated for the description of the real part of the two-time correlation function of the heat bath, $C'_{i}(t)$, and the quantity $K$ is the number of expansion coefficients for $C'_{i}(t)$.
The number of sites is represented by $N$.
The symbols $\times$ and $\circ$ represent the commutator and anticommutator defined as $\hat{O}^{\times} \bullet = [\hat{O}, \bullet]$ and $\hat{O}^{\circ} \bullet = \{\hat{O}, \bullet\}$, respectively.  We set $\nu_{i, 0} = \gamma_{i, 0}$ to simplify the notation. 
The vector $\vec{e}_{k}$ is the unit vector of the $k$th element.
To conduct numerical integrations, we set $\hat{\rho}_{\vec{n}_{1}, \ldots, \vec{n}_{N}} (t) = 0$, where $\sum_{i, k} n_{i, k} > N_{\max}$, to truncate the equations. The number $N_{\max}$ is set sufficiently large  for the calculations to converge, and it may change depending on the parameters of the system and bath.
For an outline of the derivation of the HEOM, see Appendix~\ref{sec:appHEOM}.

We truncate the eigenstates of the LO phonon in the same manner as the ADOs.
That is, the eigenstates that satisfy $\sum_{i \sigma} \hat{n}_{\sigma} > M_{LO}$ are ignored, where $M_{LO}$ is an integer representing the LO cutoff number. 
With a sufficiently large value of  $M_{LO}$, we obtain reliable results because the populations in the higher eigenstates are almost zero, owing to the rapid removal of the excess energy of the phonon modes  to the heat bath.  

\section{NUMERICAL RESULTS} \label{sec:results}
\subsection{Equilibrium distribution} \label{subsec:eqdis}
We can obtain the reduced thermal equilibrium state of the total system (which is not the Boltzmann distribution of the main system itself owing to the contribution of the system--bath interaction) by numerically integrating Eq.~\eqref{eq:HEOM} from a temporal initial state to a time sufficiently long that all of the hierarchical elements have reached a steady state.\cite{Tanimura2014,Tanimura2015,TanimuraJPSJ06, tanimuraJCP20}
 Although we can set any strength and cutoff frequency of the system--bath interaction, described as $\eta_i$ and $\gamma_i$, with any form of $\hat{V}_i$, here we limit our analysis to the weak-coupling case to justify the description of the present model in comparison with a case without a heat bath.

 We set the transfer integral, on-site Coulomb repulsion, electron--phonon coupling strength, and LO phonon frequency at each site identically as $t^i_h = t_h$, $U^i = U$, $g^i = g$, and $\Omega^i = \Omega$, respectively, for all $i$. 
 We choose the transfer integral $t_{h}$ as the unit of  frequency. Then, for all calculations, we fix the LO phonon frequency, the coupling strength between  electrons and LO phonons,  and the cutoff frequency of the heat bath as $\Omega/t_{h} = 2$, $g/t_{h} = \sqrt{3.6}$, and $\gamma_{i}/t{_h} = 3$.  
We choose these parameter values because the quantum phase transition for the SC state has been reported under such conditions.\cite{takada03hh, Tezuka07PRB}
 The inverse temperature divided by the Boltzmann constant $k_B$ is denoted by $\beta = 1 / k_B  T$.  We set $\beta t_{h} = 1$ (high-temperature case) and $\beta t_{h} = 4$ (low-temperature case).

To examine the description of the HH-plus-bath model, we calculate the expectation value of the system Hamiltonian, $\braket{\hat{H}_{S}}$, using the steady solution of the HEOM\cite{zhang17sciheom} and compare the result with the value calculated from the Boltzmann distribution of the HH system, in which the  eigenenergies are evaluated by exact diagonalization.\cite{chatterjee10hh}
For this purpose, we consider a two-site HH model with  parameter values $U/t_{h} = 3$, $M_{LO} = 5$, and $\beta t_{h} = 4$.  For the HH-plus-bath model, we consider a local heat bath described by the interaction $\hat{V}_{i} = \hat{b}^{\dagger}_{i} + \hat{b}_{i}$, in the case of  weak system--bath coupling described as $\eta_{i} = 0.001$ $(i = 1, 2)$ with the ADO parameter values $N_{\max}=2$ and $K = 2$.   We  obtain $\braket{\hat{H}_{S}} = -5.00$ for the HH-plus-bath model, which agrees very well with the value $\braket{\hat{H}_{S}} = -4.94$ obtained from  diagonalization of the HH model.  It should be noted that in the case of strong system--bath coupling, the system energy calculated from the steady-state solution of the HEOM for the HH-plus-bath model does not agree with that calculated from the Boltzmann distribution of the HH system, owing to the contribution of the energy from the system--bath interaction.

\subsection{Optical response} \label{subsec:opt}
We now consider the optical response of the HH-plus-bath system. While optical signals are, in general, formulated on the basis of the optical dipole moment, here we consider an expectation value of the optical current, $\braket{\hat{j}(t)} = \tr \{\hat{j}(t) \hat{\rho}(t)\}$, where $\hat{j}(t)=i\sum_{i, \sigma}(t_{h} e^{i(A(t) + A'(t))}\hat{c}_{i, \sigma}^{\dagger} \hat{c}_{i+1, \sigma} - \mathrm{H.c.})$ is the current operator, under a sufficiently weak probe excitation described by a time-dependent vector potential $A'(t)$. This consideration is with future extensions to bulk material in mind.  The optical conductivity is then defined as
\begin{align}
\sigma[\omega] & = \frac{1}{E'[\omega]}
\int_{0}^{\infty} dt\, e^{i \omega t}
[\braket{\hat{j}(t)} - \braket{\hat{j}(t)}_{0}  ],
\label{eq:opt}
\end{align}
where $\braket{\cdots}_{0}$ is the current without the vector potential, which implies $A'(t)=0$. 
The quantity $E'[\omega] = i \omega A'[\omega]$ is the electric field that arises from the vector potential $A'(t)$, where we denote the Fourier transform of $f(t)$ as $f[\omega] \equiv \int_{-\infty}^{\infty} dt\, e^{i \omega t} f(t)$. 
In this paper, we calculate the optical conductivity of the HH-plus-bath system for two- and four-site cases with the parameter values listed in Table~\ref{tbl:param}.

\begin{table}[!t]
\caption{System and bath parameter values for two- and four-site HH-plus-bath models.
\label{tbl:param}}
\begin{ruledtabular}
\begin{tabular}{c|c|c|c|c|c|c|c|c|c}
  & $\beta t_{h}$ & $U/t_{h}$ & $g /t_{h}$ & $\Omega / t_{h}$ & $\gamma_{i} / t_{h}$ 
  & $\eta_{i}$ & $N_{\max}$ & $K$ & $M_{LO}$ \\
  \hline 
Two-site & \multirow{2}{*}{$0.1$} & \multirow{2}{*}{$1$} & & & & $1$ & $30$ 
  & \multirow{2}{*}{$0$} & \multirow{4}{*}{$9$} \\
  \cline{7-8}
  & & & & & & \multirow{4}{*}{$0.05$} & 3 & & \\
  \cline{2-3} \cline{8-9}
  & $1$ & $1$, $3$, $5$ & $\sqrt{3.6}$ & $2$ & $3$ & & $1$ & \multirow{2}{*}{$2$} &  \\
  \cline{2-3} \cline{8-8}
  & $4$ & $1$, $3$, $5$ & & & & & $2$ & &  \\
  \cline{1-3} \cline{8-10}
  Four-site & $0.1$ & $1$ & & & & & $5$ & $0$ & $7$ 
\end{tabular}
\end{ruledtabular}
\end{table}

\subsubsection{Equilibrium response} \label{subsec:eq}
First, we consider linear absorption spectrum evaluated from Eq.~\eqref{eq:opt} with $A(t)=0$ and $E_{\mu}(t)=0$ in Eq.~\eqref{eq:H_S}. In this case, the optical conductivity is defined in terms of the linear response function as \cite{mahan00, lenarcic14oc}
\begin{align}
\label{eq:optEq}
\sigma[\omega] & = \frac{i}{\omega}\! \left (\braket{\hat{j}^{d}}_\mathrm{eq} -
\int_{0}^{\infty} dt \, e^{i \omega t} R^{(1)}(t) \right),
\end{align}
where 
\begin{align}
R^{(1)}(t)&= i\theta(t)\braket{[\hat{j}^{p}(t), \hat{j}^{p}(0)]}_\mathrm{eq}, \\[6pt]
 \hat{j}^{d} &= t_{h}\sum_{i, \sigma} (\hat{c}_{i, \sigma}^{\dagger} \hat{c}_{i+1, \sigma} + \mathrm{H.c.}), \\[3pt]
  \hat{j}^{p} &= it_{h} \sum_{i, \sigma} (\hat{c}_{i, \sigma}^{\dagger} \hat{c}_{i+1, \sigma} - \mathrm{H.c.})
\end{align}
 are the linear response function and the diamagnetic and paramagnetic current operators, respectively.\cite{bergeron11oc} Here,  $\theta(t)$ is the step function.
For details of the derivation of Eq.~\eqref{eq:optEq}, see 
Appendix~\ref{sec:appOpt}.

To compare the present results with those obtained previously from the HH model without a heat bath, we consider a fast modulation case with a weak system--bath interaction described as $\gamma_{i}/t_{h} = 3$ and  $\eta_{i} = 0.05$ $(i=1, 2)$. We then chose the on-site Coulomb repulsion as $U/t_{h} = 1$, $3$, and $5$ and the inverse temperature as $\beta t_{h} = 1$ and $4$.  

\begin{figure}[!t]
\includegraphics[width=\linewidth]{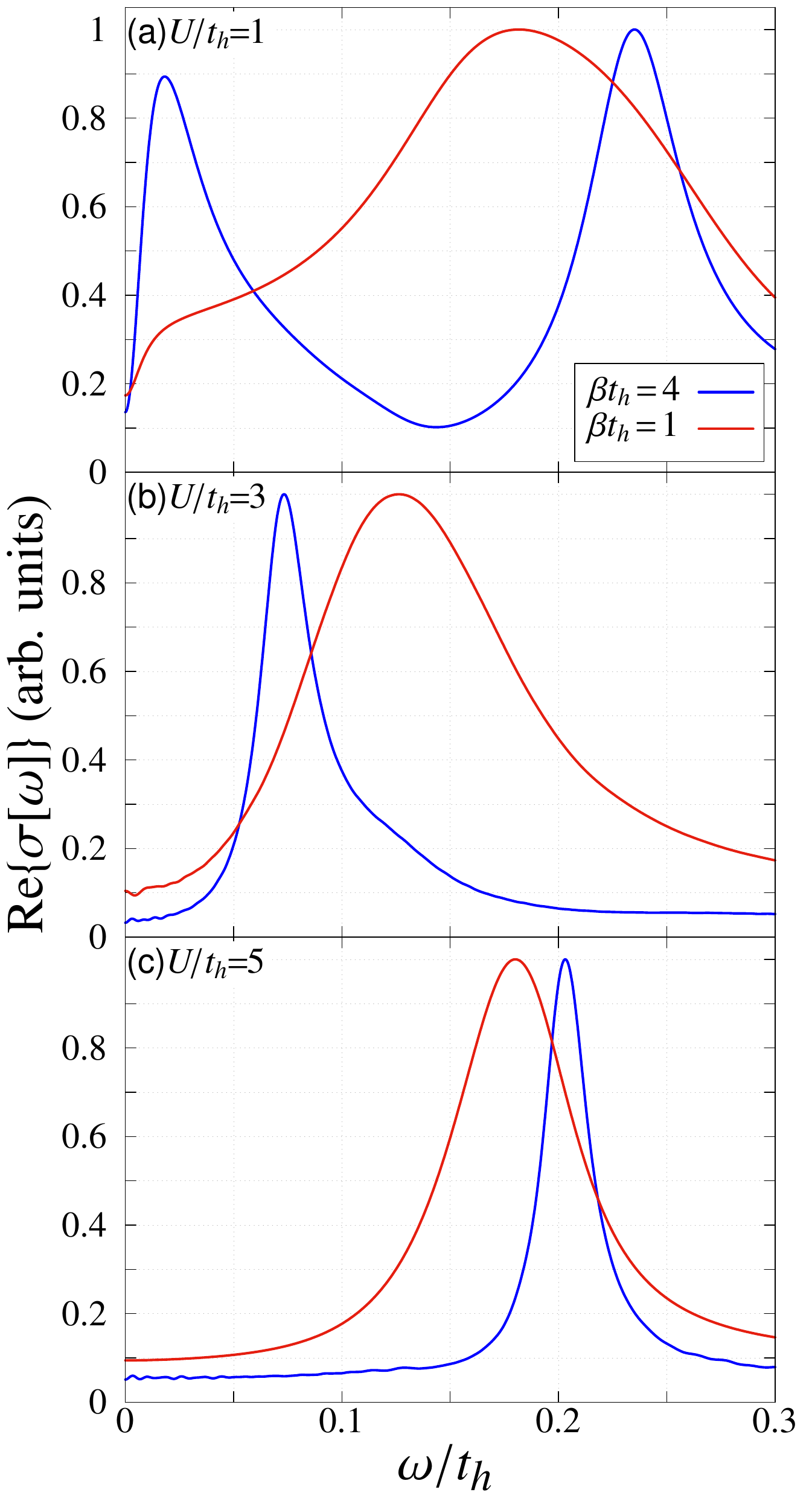}
\caption{Real part of the optical conductivity $\Re \{\sigma[\omega]\}$ for (a)  weak  ($U/t_{h} = 1$), (b)  intermediate  ($U/t_{h} = 3$), and (c) strong ($U/t_{h} = 5$)  Coulomb repulsion in  the low-temperature case $\beta t_{h} = 4$ (blue curves) and the high-temperature case $\beta t_{h} = 1$ (red curves)  calculated from Eq.~\eqref{eq:optEq}. 
The LO phonon frequency and the electron--phonon coupling strength are chosen as $\Omega /t_{h} = 2$ and $g/t_{h} = \sqrt{3.6}$, respectively.
Each line shape is normalized such that the maximum peak in the region $0 \leq \omega \leq 0.3$ is $1$.  The entire peak profiles are presented in Appendix~\ref{sec:appPeaks}.
\label{fig:OcEq}}
\end{figure}

It should be mentioned that, in a case of a bulk HH system without a heat bath, we observe continuous energy transition bands. If the bulk system is a conductor, we observe the peak at zero frequency. In the present two-site model, however, continuous bands do not appear, owing to the finite number of  eigenstates, and the peak representing the conducting state appears at nonzero frequency.

Figure~\ref{fig:OcEq} displays the low-frequency region of the real part of the optical conductivity, $\Re \{\sigma [\omega] \}$, for the two-site HH-plus-bath model. The peaks over the entire spectral region are presented in Appendix~\ref{sec:appPeaks}.
In the low-temperature case  $\beta t_{h} = 4$ (blue curves), the electronic transition peaks are observed at $0.235, 0.073$, and $0.202$ for (a)  weak ($U/t_{h} = 1$), (b) intermediate ($U/t_{h} = 3$), and (c) strong ($U/t_{h} = 5$) Coulomb repulsion, respectively. In the weak-repulsion case in Fig.~\ref{fig:OcEq}(a), we also observe a peak at $0.019$, which is due to the Drude weight reflecting the sensitivity of the system to an external vector potential. This peak appears even in the case of an insulator.\cite{castella95integrability, zotos96integrability} 
The appearance of this peak depends strongly  on the system size, and hence the peak disappears as the system size increases.  Therefore,  we shall not discuss this low-frequency peak further here, and we focus our analysis only on higher transition peaks.   
As Figs.~\ref{fig:OcEq}(a)--\ref{fig:OcEq}(c) indicate, the position of the transition peak first decreases then increases as a function of $U$. This tendency agrees with the previous results obtained on the basis of the HH model, as presented in Ref.~\onlinecite{Tezuka07PRB} (note that the present parameter value $g = \sqrt{3.6}$ is equal to the value $\lambda = 3.6$ in that paper). This implies that the weak-repulsion case corresponds to a charge-ordered insulator, whereas the strong-repulsion case corresponds to an AF insulator.

In the high-temperature case ($\beta t_{h} = 1$), the peak position for intermediate Coulomb repulsion  in Fig.~\ref{fig:OcEq}(b) is blue-shifted, indicating that the insulator-like behavior is enhanced. By contrast, the peak positions in  the small- and large-$U$ cases in Figs.~\ref{fig:OcEq}(a) and \ref{fig:OcEq}(c), respectively, are red-shifted. Although the present system is not yet a conductor, the blue-shifted peak in the intermediate-repulsion case does suggest the presence of the conductor--insulator phase transition that is observed in a bulk HH system.

\subsubsection{Effects of different forms of system--bath coupling  and of system size}
Above, we considered the situation in which each LO phonon mode is coupled to its own bath. Moreover, we assumed that the system part of the system--bath interaction was a linear function of phonon coordinate, as $\hat{V}_{i}= \hat{b}_{i}^{\dagger} + \hat{b}_{i} \propto \hat{q}_i$, which leads to population relaxation and excitation.\cite{TanimuraJPSJ06} This form of interaction has been used in studies of vibrational spectroscopy and electron transfer  on the basis of a Brownian oscillator model.\cite{TanimuraJPSJ06,tanimuraJCP20,Onuchic,Ueno2021,TaniMukaJPSJ90,TanakaHEOM1,TanakaHEOM2,TJCP2012,Mauro2021}  When the vibrational excitation energy becomes close to the thermal excitation energy, this process becomes important.  If the vibrational excitation energy is much higher than the thermal excitation energy, however,   vibrational dephasing described by $\hat{V}_{i} \propto \hat{q}_{i}^2$ plays a significant role. \cite{OkumuraPRE1997, IshizakiJCP2006}  Under such conditions, this interaction can be approximated in  diagonal modulation form as $\hat{V}_{i} \approx \hat{b}^{\dagger}_{i} \hat{b}_{i}$.\cite{TanimuraJPSJ06,tanimuraJCP20} In this subsection, we compare the different forms of  interactions  for vibrational dephasing ($\hat{V}_{i} = \hat{b}^{\dagger}_{i} \hat{b}_{i}$) and  vibrational relaxation ($\hat{V}_{i} = \hat{b}^{\dagger}_{i} + \hat{b}_{i}$).
Moreover, we examine the difference between the local and nonlocal ($\hat{V}_{0} = \sum_{i} \hat{b}^{\dagger}_{i} \hat{b}_{i}$) system--bath interactions, as well as the  effects of  system size.

In the following, we  consider only a high-temperature case ($\beta t_{h} = 0.1$) and set $K=0$ and $c_{i, 0} = \eta_{i} \gamma_{i} / \beta$ to reduce the numerical costs of integrating the HEOM. 
In the present case, owing to the strong relaxation effects, it is not easy to apply Eq.~\eqref{eq:opt} to calculate the optical conductivity. Therefore, we evaluate this from Eq.~\eqref{eq:optEq}.

\begin{figure}[!t]
\includegraphics[width=\linewidth]{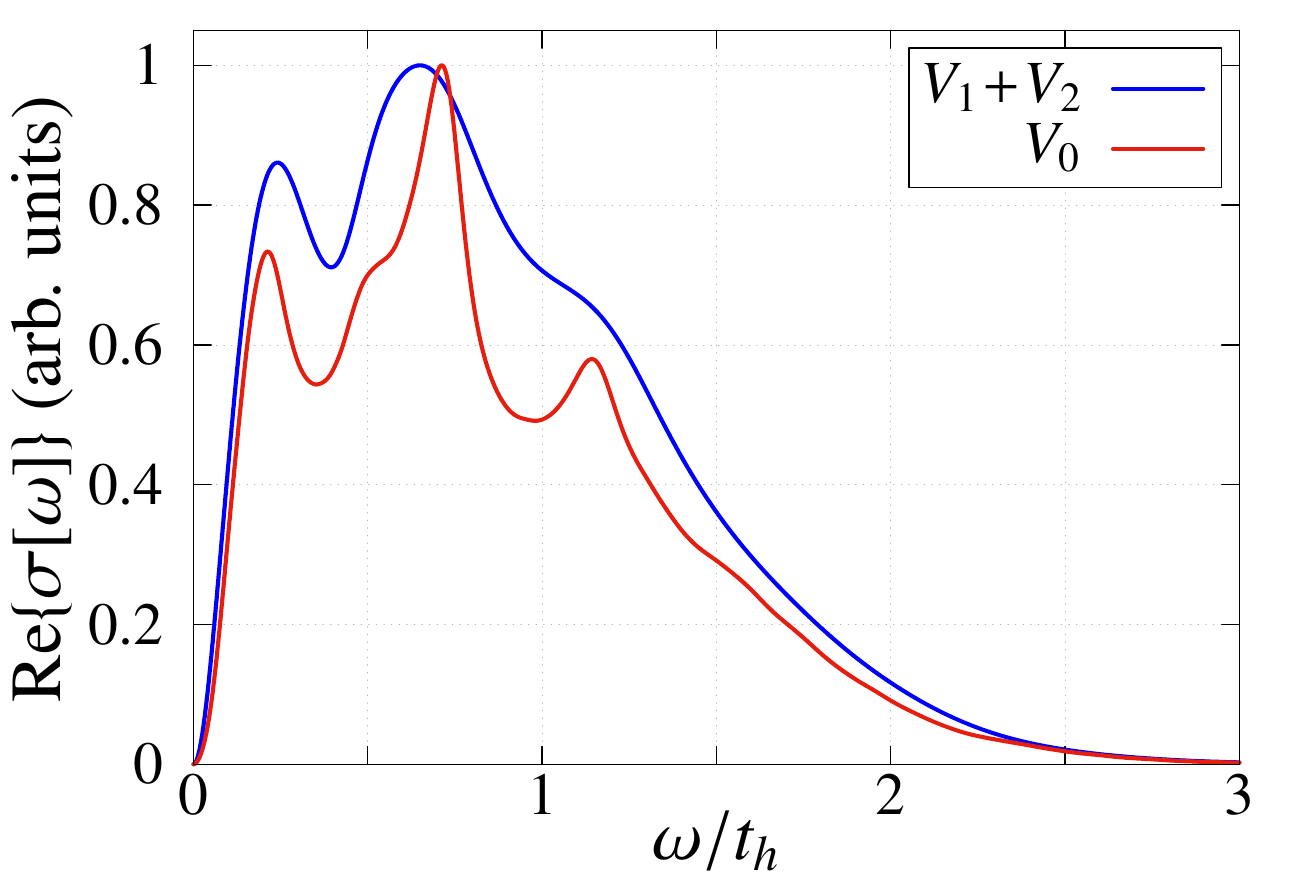}
\caption{
Optical conductivity $\Re \{\sigma(\omega]\}$ in the local vibrational-relaxation case, $\hat{V}_{i} = \hat{b}^{\dagger}_{i} + \hat{b}_{i}$, in which the LO mode of each site is modulated independently (blue curve), and that in the nonlocal bath case, $\hat{V}_{0} = \sum_{i} \hat{b}^{\dagger}_{i} \hat{b}_{i}$, in which both of the LO modes are coupled to a single bath (red curve). The Coulomb repulsion and inverse temperature are set as $U/t_h = 1$ and $\beta t_h = 0.1$, and the other parameter values are the same as in Fig.~\ref{fig:OcEq}, except  those for the ADOs.
\label{fig:VComp}
}
\end{figure}

In Fig.~\ref{fig:VComp}, we present the calculated results of $\Re \{\sigma[\omega]\}$ for the independent-local (blue curve) and nonlocal (red curve) bath coupling cases.  The peaks in the nonlocal bath case are narrower and not shifted in comparison with the peaks in the local bath case, because the interaction $\hat{V}_{0} = \sum_{i} \hat{b}^{\dagger}_{i} \hat{b}_{i}$  contributes  only indirectly to  energy relaxation, and  the nonlocal heat bath enhances the intersite quantum coherence among the phonon modes.

\begin{figure}[!t]
\includegraphics[width=\linewidth]{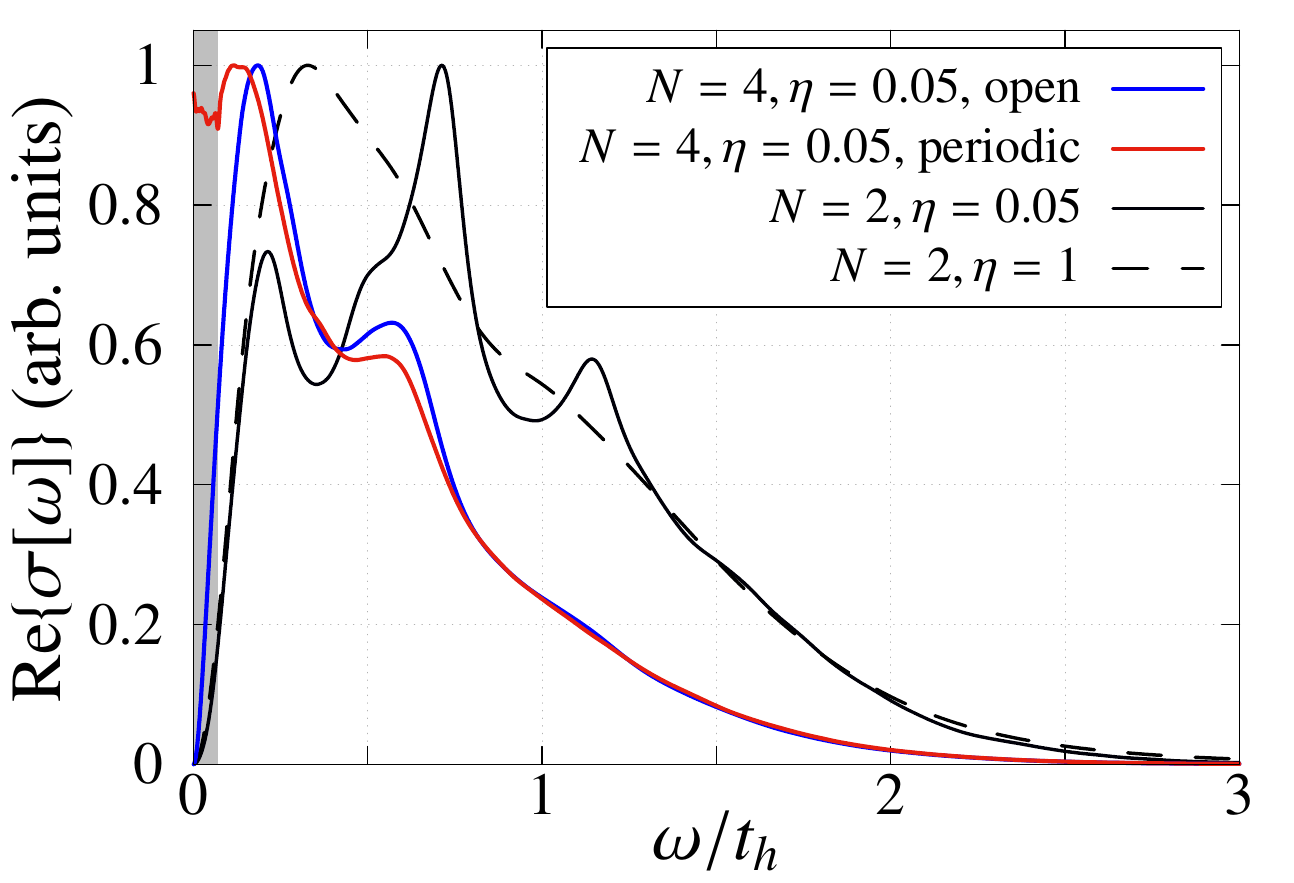}
\caption{
Optical conductivity $\Re \{\sigma[\omega]\}$ for diagonal system--bath coupling with $\eta=0.05$  for the two-site ($N = 2$) case (solid black curve) and for  four-site ($N = 4$) cases with open (blue curve) and periodic (red curve) boundary conditions. For the purpose of illustration, we also depict the two-site case with  stronger coupling $\eta=1$ (dashed black curve). The Coulomb repulsion and inverse temperature are set as $U/t_h = 1$ and $\beta t_h = 0.1$. The other parameter values are the same as in Fig.~\ref{fig:OcEq}, except  those for the ADOs. Note that owing to  numerical error (marked in the gray area), the four-site result with  periodic  boundary condition does not approach  zero at zero frequency. 
\label{fig:N2-4}
}
\end{figure}

Next, we discuss the  effects of  system size. To reduce the computational cost,  we consider the nonlocal bath case only. In Fig.~\ref{fig:N2-4}, the blue and red curves represent the results with $\eta=0.05$ in  four-site cases with  open and periodic boundary conditions, respectively.  We set $N_{\max} = 5$ and $M_{LO} = 7$, with the other parameter values being the same as in Fig.~\ref{fig:VComp}.  Note that we set slightly smaller values of $M_{LO}$ to reduce the computational cost for the investigation of  size effects; we have verified that the line shapes for $M_{LO} = 9$ and $7$ are qualitatively the same in the two-site case (results are not shown).  For reference purposes,  we also re-plot here as the solid black curve the two-site result depicted in Fig.~\ref{fig:VComp}. As these results indicate, the number of peaks is suppressed for a larger system. This is because we consider the system part of the system--bath coupling in the form $\hat{V}_{0} = \sum_{i}^N \hat{b}^{\dagger}_{i} \hat{b}_{i}$, and the number of system--bath couplings, which determines the effective coupling strength for the system, increases as the number of  sites increases. 

To confirm this, we depict the two-site result for the case of large system--bath coupling  ($\eta=1$) as the dashed black curve in Fig.~\ref{fig:N2-4}; as $\eta$ increases, the peaks in the high-frequency region are suppressed, while those in the low-frequency region are merged and enhanced. 
As the number of sites increases, the energy gap decreases because of the increased number of  eigenstates, which is reminiscent of  band theory. Thus, the lowest peak in the four-site case is red-shifted in comparison with that in the two-site case.  This tendency is more pronounced for the periodic boundary condition than for the open boundary condition, because the effective size of the system is larger in the former case. 

\begin{figure}[h]
  \includegraphics[width=\linewidth]{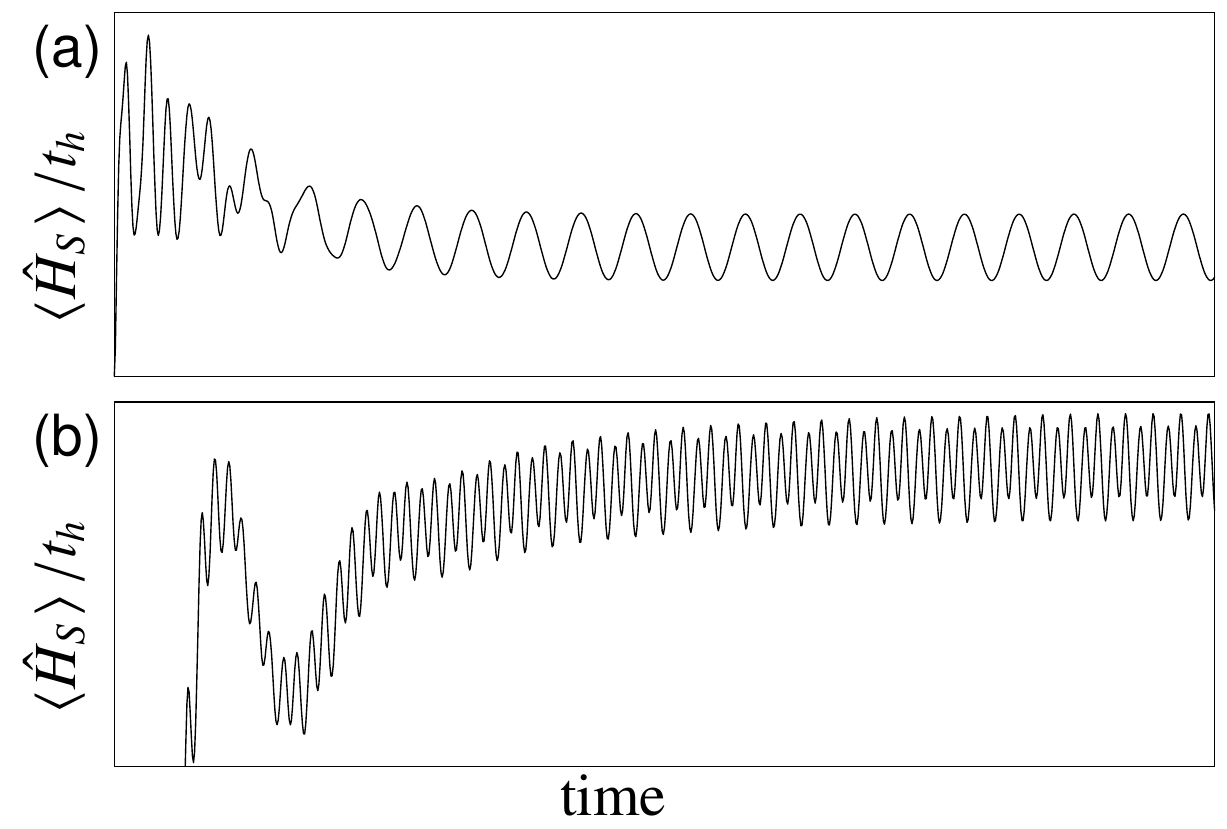}
  \caption{Time evolution of the expectation value of $\hat{H}_{S}$ under external electric fields for (a) electric excitation and (b) LO phonon excitation.
  \label{fig:EneNeq}}
\end{figure}
\subsubsection{Continuous wave response} \label{subsec:noneq}
Finally, we discuss the steady-state response under periodic external electric fields. Here, we consider the case of off-diagonal system--bath coupling  ($\hat{V}_{i} = \hat{b}_{i}^{\dagger} + \hat{b}_{i}$) for (a) the electronic excitation
\begin{align}
\begin{aligned}
A(t) &=  A_{\mathrm{ex}} \cos (\Omega_{\mathrm{ex}}t), \\[3pt]
 E_{\mu} (t) &= 0,
\end{aligned}
  \label{eq:activeE}
\end{align}
and  (b) the LO phonon excitation,
\begin{align}
\begin{aligned}
A(t) &=   0, \\[3pt]
E_{\mu} (t) &= \mu E_{\mathrm{ex}} \cos (\Omega_{\mathrm{ex}} t ).
\end{aligned}
  \label{eq:activePh}
\end{align}
Under an external time-dependent perturbation, we cannot evaluate the optical conductivity from Eq.~\eqref{eq:optEq}, because there is no thermal equilibrium state in this situation.  We therefore simulate Eq.~\eqref{eq:opt} explicitly.
For an external field $A'(t) = A_{\mathrm{probe}}\theta(t-t_{\mathrm{on}})$, Eq.~\eqref{eq:opt} is expressed as (see Appendix~\ref{sec:appOpt}):
\begin{align}
\sigma [\omega] = - \frac{1}{A_{\mathrm{probe}}}
 \int_{0}^{\infty} dt\, e^{i \omega (t-t_{\mathrm{on}})}
[\braket{\hat{j}(t)} - \braket{\hat{j}(t)}_{0}].
\label{eq:optSimple}
\end{align}
For sufficiently small $A_{\mathrm{probe}}$ (from $5\times 10^{-3}$ to $1 \times 10^{-2}$), the optical conductivity defined by Eq.~\eqref{eq:opt} agrees with that defined by Eq.~\eqref{eq:optEq} in the equilibrium case.\cite{shao16oc}
Because of the prefactor $e^{-i \omega t_{\mathrm{on}}}$, the profile of the optical conductivity changes depending on $t_{\mathrm{on}}$, but that dependence is minor anyway, and so we ignore it in the following discussion.

Here, we consider the high-temperature case ($\beta t_{h} = 1$) in order to find signs of the photoinduced phase transition from insulator to conductor that is observed in bulk systems. When the HEOM are integrated over a sufficiently long time, the system driven by the periodic external fields in Eqs.~\eqref{eq:activeE} and \eqref{eq:activePh} reaches a time-dependent steady state. The transient behaviors as characterized by the system energy are depicted in Fig.~\ref{fig:EneNeq}. In the present theory of  open quantum dynamics, the energy supplied by the external fields dissipates to the heat bath, whose time-irreversible behavior is characterized by the fluctuation--dissipation theorem, whereas a wavefunction-based Schr\"odinger approach cannot properly treat thermal effects characterized by a temperature.

\begin{figure*}[!t]
  \includegraphics[width=\linewidth]{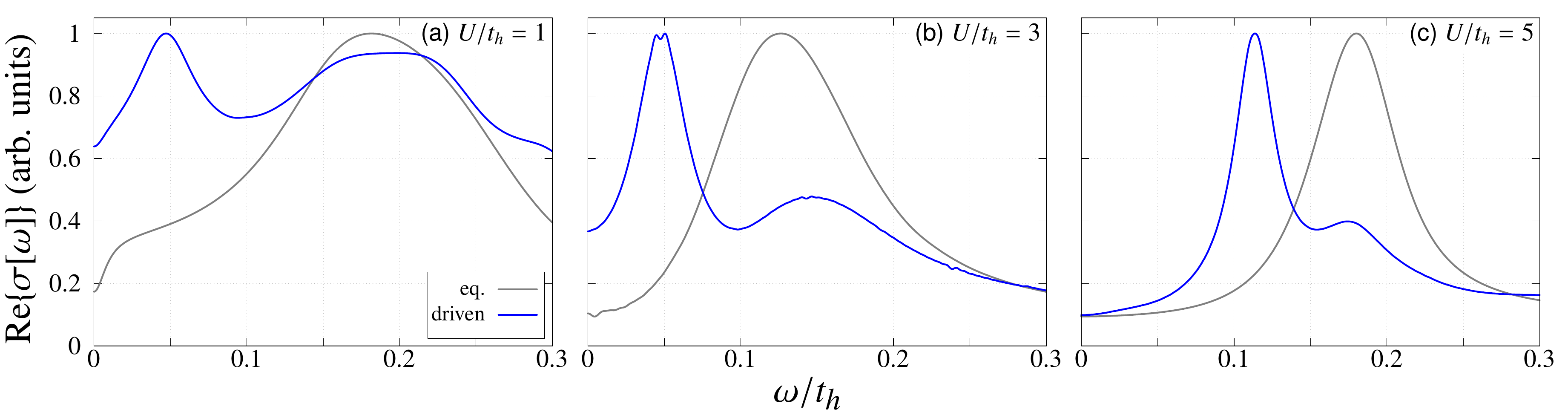}
  \caption{Optical conductivity $\Re \{\sigma[\omega]\}$ under the electronic excitation presented in Eq.~\eqref{eq:activeE} for  (a) weak ($U/t_h = 1$),  (b) intermediate ($U/t_h = 3$), and (c) strong ($U/t_h = 5$) Coulomb repulsion. 
  The gray curves represent the optical conductivity of the equilibrium states at  high temperature (the same results as shown by the red curve in Fig.~\ref{fig:OcEq}).
  The amplitude $A_{\mathrm{ex}}$ is set to $0.5$.
  \label{fig:OcNeqE}}
\end{figure*}

\begin{figure*}[!t]
  \includegraphics[width=\linewidth]{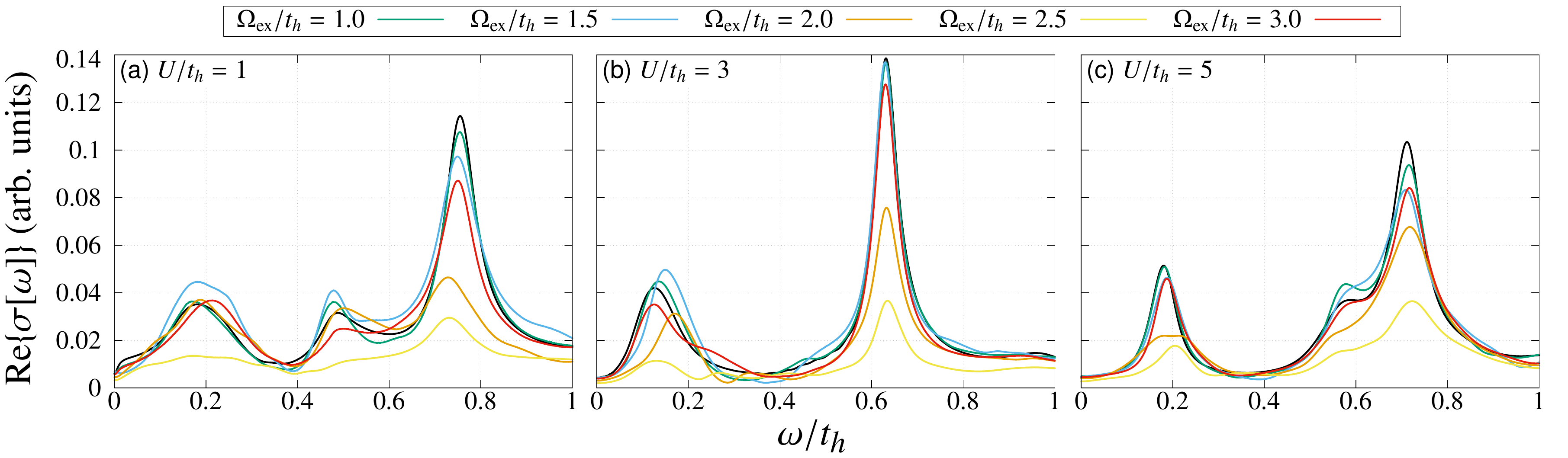}
  \caption{Optical conductivity $\Re \{\sigma[\omega]\}$ under the LO phonon excitation presented in Eq.~\eqref{eq:activePh}  for various values of the excitation frequency $\Omega_{\mathrm{ex}}$ for  (a) weak ($U/t_h = 1$),  (b) intermediate ($U/t_h = 3$), and (c) strong ($U/t_h = 5$) Coulomb repulsion.
  The inverse temperature is set as $\beta t_{h} = 1$.
  The black lines represent the optical conductivity of the equilibrium states at  high temperature (the same results as shown by the red curve  in Fig.~\ref{fig:OcEq}).
 None of the results are normalized. The amplitude $\mu E_{\mathrm{ex}}$ is set to $0.5$.
  \label{fig:OcNeqPh}
  }
\end{figure*}

In Fig.~\ref{fig:OcNeqE}, we present the calculated optical conductivity under the electronic excitation given by Eq.~\eqref{eq:activeE}. 
For a bulk Hubbard system, it has been reported that by tuning the frequency of the external electronic field to the magnitude of Coulomb repulsion $U$, the insulating Hubbard model becomes superconducting.\cite{Kaneko19PRL}
Here, we examine this behavior using the HH-plus-bath model. 

For this purpose, we have to estimate the effective strength of the Coulomb repulsion in the HH-plus-bath model.  In contrast to the Hubbard model, this estimation is difficult for both the HH and HH-plus-bath cases, because the transition energies are altered from the Hubbard case owing to the contribution from  electron--phonon coupling.  In the strong-repulsion case, however, the electron--phonon contribution is minor, and we find that the effective Coulomb repulsion can be evaluated from the peak position of the optical conductivity, in the same way as in the two-site Hubbard model (see Appendix~\ref{sec:appPeaks}).  

Therefore, in Fig.~\ref{fig:OcNeqE}(c)   in the strong-repulsion case ($U/t_h = 5$), we  apply an external electric field with  frequency $\Omega_{\mathrm{ex}} / t_h =  0.71$ and find that the peak shifts to the red by approximately $0.11$, which indicates that the driven system is in the conducting state.  This peak shift can be attributed to   $\eta$ pairing, as in the Hubbard case.\cite{Kaneko19PRL}

In the intermediate-repulsion ($U / t_h = 3$) and weak-repulsion ($U  / t_h = 1$) cases, the frequency of the main peak in $\Re \{\sigma[\omega]\}$ does not agree with that given by the effective Coulomb repulsion, because of the contribution from the LO phonon interaction. Thus, the peak does not shift, even when the frequency of the external field is tuned to the dominant peak position, with $\Omega_{\mathrm{ex}} / t_h = 0.75$ (weak-repulsion case) and $\Omega_{\mathrm{ex}} / t_h = 0.63$ (intermediate-repulsion case).  We then vary the resonant frequency $\Omega_{\mathrm{ex}}$ to find out whether $\eta$ pairing occurs in the weak- and intermediate-repulsion cases, and we set $\Omega_{\mathrm{ex}} / t_h = 0.3$.  We  display $\Re \{\sigma[\omega]\}$ for the  weak- and intermediate-repulsion cases in Figs.~\ref{fig:OcNeqE}(a) and  \ref{fig:OcNeqE}(b), respectively.
In the intermediate-repulsion case, the peak is red-shifted from $0.13$ to $0.05$, while the peak profile is similar to that in the strong-repulsion case: We find that  $\eta$ pairing occurs in this situation, although  estimation of the effective Coulomb repulsion remains difficult.
In the weak-repulsion case, although a peak shift can be observed, the profiles of the two peaks at $0.05$ and $0.18$ are different from those in Figs.~\ref{fig:OcNeqE}(b)  and  \ref{fig:OcNeqE}(c). This indicates the suppression of  $\eta$ pairing.

Although our system is small,  we may conclude here that  $\eta$ pairing, which plays a key role in the superconductivity of the Hubbard system, is suppressed when the electron--phonon coupling becomes stronger than the Coulomb repulsion, while  superconductivity may appear when the Coulomb repulsion is stronger or comparable to the electron--phonon coupling, even when the HH system is coupled to a heat bath.

In Fig.~\ref{fig:OcNeqPh}, we present the calculated optical conductivity under the LO phonon excitation given by Eq.~\eqref{eq:activePh} for various values of the laser frequency $\Omega_{\mathrm{ex}}$ and Coulomb repulsion $U$. 
 In all three cases with different strengths of repulsion  $U$, the position of the lowest peak near $\omega/t_h=0.2$ does not change very much as a function of $\Omega_\mathrm{ex}/t_h$, while the peak intensities are suppressed for $\Omega_{\mathrm{ex}} / t_{h} =2.5$, which is approximately the same as the effective resonant frequency of the LO phonon mode under  off-diagonal system--bath coupling. In this case, the LO phonon mode is strongly excited by the external electric field. As a result, a polaron is created, while  electron transfer is suppressed. This indicates that the LO phonon excitation does not lead to  photoinduced superconductivity. 

To observe superconductivity with photoexcited phonons, which has been reported in previous studies, it may be necessary to include Coulomb repulsion between  nearest-neighbor sites \cite{matsueda11exhh, hashimoto17ih, seo18h} and to change the distance between molecules.\cite{yonemitsu12exphh}

\section{CONCLUDING REMARKS} \label{sec:conclude}
In this paper, we introduced the HH-plus-bath model to investigate the effects of a short-range repulsive Coulomb interaction and a local optical vibrational mode for the investigation of  electron dynamics under nonequilibrium conditions. We employed the HEOM formalism to simulate the dynamics of electrons described by this model under various physical conditions for local and nonlocal LO phonon--bath interactions in a numerically rigorous manner. We calculated the optical conductivity at finite temperature by integrating the HEOM and found that the present model exhibits conductor-like behavior when the Coulomb repulsion and the coupling strength between the electrons and the LO phonons are comparable. This finding agrees with  previous results obtained for an HH bulk system. 

We then calculated the optical properties of the system under continuous-wave laser excitation. 
We found that the excitation of the electrons promotes  photoinduced conductivity when the Coulomb repulsion is comparable to or larger than the electron--phonon interaction, while the electrical properties do not change in the region in which  electron--phonon coupling plays a role. We found that  excitation of  LO phonons always suppresses the conductivity. Although  we restricted our analysis here to the case of a steady-state response, there is no inherent restriction on the dynamical simulation  with use of the HEOM. Thus, it is straightforward to compute ultrafast nonlinear spectra, including two-dimensional electronic--vibrational spectra, on the basis of the present formulation.

In the present work, we limited our analysis to a small system with a specific parameter set focusing on the role of local electron--phonon interactions. Because  electron transfer is a long-range effect, an extension of the present investigation to a larger system would  provide deeper insight, in particular, with regard to studies of  superconductivity. Thus, to make the present approach more useful, further computational efforts need to be made to treat larger systems that consist of many sites, for example, by employing the hierarchical Schr\"{o}dinger equations of motion \cite{Nakamura18PRA} and the tensor-train method \cite{BorrelliTT19, BorrelliTT21, ShiTT18, ShiTT21}. Such investigations are left for future work. Nevertheless, we believe that the present results elucidate the key features of the HH model under thermal conditions with regard to the fundamental nature of  strongly correlated electron systems.

\begin{acknowledgments}
The calculations reported here are supported in part by the Research Center for Computational Science, National Institute of Natural Sciences.
Y.T. is supported by JSPS KAKENHI Grant No. B 21H01884.
\end{acknowledgments}

\section*{Data availability}
The data that support the findings of this study are available from the corresponding author upon reasonable request.

\appendix

\section{DERIVATION OF THE HEOM FOR THE HH-PLUS-BATH SYSTEM} \label{sec:appHEOM}
In this appendix, we construct the HEOM for the HH-plus-bath system.
The reduced density operator of the system, $\hat{\rho}(t) = \tr_{B} \{e^{-i \hat{H}_\mathrm{tot}t} \hat{\rho}_\mathrm{tot}(0) e^{i \hat{H}_\mathrm{tot}t}\}$, is expressed in the coherent state representation of the path integral as
\begin{widetext}
\begin{multline}
\label{eq:RDO}
 \braket{\bm{\xi}, \bm{\alpha}| \hat{\rho} (t)| \bm{\xi}', \bm{\alpha}'} = \\
 \int \mathcal{D}^{2} \bm{\xi}\, \mathcal{D}^{2} \bm{\xi}' \int  \mathcal{D}^{2} \bm{\alpha}\, \mathcal{D}^{2} \bm{\alpha}' \int d^{2}\bm{\xi}_{i}  d^{2}\bm{\xi}'_{i} \int  \frac{d^{2}\bm{\alpha}_{i}}{\pi^{N}} \frac{d^{2}\bm{\alpha}'_{i}}{\pi^{N}}  
    e^{i S_{S}(\bm{\xi}, \bm{\alpha}; t)} \braket{\bm{\xi}_{i}, \bm{\alpha}_{i}|\hat{\rho}(0)|\bm{\xi}'_{i}, \bm{\alpha}'_{i}} e^{-i S_{S}(\bm{\xi}', \bm{\alpha}'; t)}  
    \mathcal{F}(\bm{\alpha}, \bm{\alpha}'; t).
\end{multline}
Here, the electronic states are described as a set of Grassmann numbers $\bm{\xi} = (\xi_{1, \uparrow}, \xi_{1, \downarrow}, \xi_{2, \uparrow}, \ldots, \xi_{N, \downarrow})$, and $\ket{\bm{\xi}}$ is the eigenstate of fermions (electrons) in the coherent state representation. The LO phonons are described as a set of complex numbers $\bm{\alpha} = (\alpha_{1}, \alpha_{2}, \ldots, \alpha_{N})$, and $\ket{\bm{\alpha}}$ are the eigenstates of bosons (LO phonons) in the coherent state representation.
The functional $S_{S}(\bm{\xi}, \bm{\alpha}; t)$ is the action of the system and is given by
\begin{align}
 S_{S}(\bm{\xi}, \bm{\alpha}; t)
 = \int_{0}^{t} dt'
\left[i\big\{\bm{\xi}^{*}(t') \cdot \dot{\bm{\xi}}(t') 
  + \tfrac{1}{2}\big[\bm{\alpha}^{*}(t') \cdot \dot{\bm{\alpha}}(t') - \dot{\bm{\alpha}}^{*}(t') \cdot \bm{\alpha}(t')\big] \big\} 
  -  \left(H_{S}(\bm{\xi}, \bm{\alpha};t') + \sum_{i} \lambda_{i} V_{i}^{2}(\alpha_{i};t') \right) \right],
\end{align}
and $H_{S}(\bm{\xi}, \bm{\alpha};t')$ and $V_{i}(\alpha_{i};t)$ are the path-integral representations of $\hat{H}_{S}$ and $\hat{V}_{i}$. The influence functional $\mathcal{F}(\bm{\alpha}, \bm{\alpha}'; t)$ is evaluated as\cite{Tanimura2014,Tanimura2015,TanimuraJPSJ06}
\begin{align}
   \mathcal{F}(\bm{\alpha}, \bm{\alpha}'; t) 
  =  \prod_{i} \exp\!
   \left(-\int_{0}^{t} dt' \int_{0}^{t'} dt''\,
  V_{i}^{\times}(\alpha_{i}, \alpha_{i}';t')  
  \!\left[C'_{i}(t'-t'')
  V_{i}^{\times}(\alpha_{i}, \alpha_{i}'; t'') 
    -i C''_{i}(t'-t'') V_{i}^{\circ}(\alpha_{i}, \alpha_{i}'; t'') \right]
\right).
\end{align}
\end{widetext}
The hyper-operators are defined as $V_{i}^{\times}(\alpha_{i},\alpha_{i}';t) \equiv V_{i}(\alpha_{i};t) - V_{i}(\alpha_{i}';t)$ and $V_{i}^{\circ}(\alpha_{i}, \alpha_{i}';t) \equiv V_{i}(\alpha_{i};t) + V_{i}(\alpha_{i}';t)$.
The real and imaginary parts of the bath correlation functions of the $i$th heat bath, $C'_{i}(t)$ and $C''_{i}(t)$, are expressed using the SDF as
\begin{align}
  C'_{i}(t) &= \int_{0}^{\infty} d \omega \,J_{i}(\omega) \coth\!\left(\frac{\beta \omega}{2}\right) \cos \omega t,
\\[6pt]
  C''_{i}(t) &= \int_{0}^{\infty} d\omega \,J_{i}(\omega) \sin \omega t,
\end{align}
where $\beta$ is the inverse temperature divided by the Boltzmann constant, $\beta =1/k_B T$.
By using the Drude--Lorentz SDF in Eq.~\eqref{eq:sdf}, the reorganization energy and the two-time correlation function of the heat bath are characterized by the parameters $\eta_{i}$ and $\gamma_{i}$, as $\lambda_{i} = \eta_{i} \gamma_{i}/2$, $C'_{i}(t) = c_{i, 0}e^{-\gamma_{i}|t|} + \sum_{k} c_{i, k} e^{-\nu_{i, k}|t|}$, and $C''_{i}(t) = \mathrm{sgn}(t)\eta_{i} \gamma_{i}^{2} e^{-\gamma_{i} |t|} /2$.
The parameters $\{c_{i, k}\}$ and $\{\nu_{i, k}\}$ are determined by a Matsubara spectral decomposition scheme \cite{heom} and a Pad{\' e} spectral decomposition scheme.\cite{YanPade11}
In this paper, we use the $[N-1/N]$ Pad{\'e} spectral decomposition scheme.

To obtain differential equations in time, we  consider the time derivative defined as
\begin{widetext}
\begin{gather}
\frac{\partial}{\partial t} \hat{\rho} (t) \equiv
\int d^2 \bm{\xi}\, d^2 \bm{\xi}' \int \frac{d^2 \bm{\alpha}}{\pi ^N} \frac{d^2 \bm{\alpha}'}{\pi ^N}
\ket{\bm{\xi}, \bm{\alpha}}
\lim_{\Delta t \to 0} \frac{\braket{\bm{\xi}, \bm{\alpha}| \hat{\rho} (t+\Delta t)| \bm{\xi}', \bm{\alpha}'}
- \braket{\bm{\xi}, \bm{\alpha}| \hat{\rho} (t)| \bm{\xi}', \bm{\alpha}'}} {\Delta t}
\bra{\bm{\xi}', \bm{\alpha}'}.
\label{eq:timeDerivative}
\end{gather}
The details of the evaluation of the right-hand side of Eq.~\eqref{eq:timeDerivative} are described in Refs.~\onlinecite{heomHigh, TanimuraJPSJ06, tanimuraJCP20,Tanimura2014,Tanimura2015}. The auxiliary density operators (ADOs) presented in Eq.~\eqref{eq:HEOM} are defined as
\begin{align}
\label{eq:ADO}
 \braket{\bm{\xi}, \bm{\alpha}| \hat{\rho} _{\vec{n}_{1}, \ldots, \vec{n}_{N}} (t)| \bm{\xi}', \bm{\alpha}'} 
 ={}& \int \mathcal{D}^{2} \bm{\xi}\, \mathcal{D}^{2} \bm{\xi}' \int \mathcal{D}^{2} \bm{\alpha} \,\mathcal{D}^{2} \bm{\alpha}' \int d^{2}\bm{\xi}_{i}\, d^{2}\bm{\xi}'_{i} \int \frac{d^{2}\bm{\alpha}_{i}}{\pi^{N}} \frac{d^{2}\bm{\alpha}'_{i}}{\pi^{N}} \nonumber\\[3pt]
   &\times \prod_{i=1}^{N}
    \left\{ \left[\int_{0}^{t} dt'' e^{-\nu_{i, 0}(t-t'')}
  \left(-i c_{i, 0} V_{i}^{\times}(\alpha_{i}, \alpha_{i}';t'')
   -\frac{\eta_{i}\nu_{i, 0}^{2}}{2} V^{\circ}(\alpha_{i}, \alpha_{i}';t'')\right) \right]^{n_{i, 0}} \right. 
 \nonumber \\[3pt]
  & \left. \times \prod _{k=0}^{K} \left(\int_{0}^{t} dt''\, e^{-\nu_{i, k}(t-t'')}\big[-i c_{i, k} V_{i}^{\times}(\alpha_{i}, \alpha_{i};t'')\big]\right)^{n_{i, k}} \right\} \nonumber \\[3pt]
  & \times e^{i S_{S}(\bm{\xi}, \bm{\alpha}; t)} \braket{\bm{\xi}_{i}, \bm{\alpha}_{i}|\hat{\rho}(0)|\bm{\xi}'_{i}, \bm{\alpha}'_{i}} e^{-i S_{S}(\bm{\xi}', \bm{\alpha}'; t)} 
  \mathcal{F}(\bm{\alpha}, \bm{\alpha}'; t),
\end{align}
in which the zeroth-order term $\hat{\rho}_{\vec{0}, \ldots, \vec{0}}(t)$ corresponds to Eq.~\eqref{eq:RDO}.
\end{widetext}

\begin{figure*}
  \includegraphics[width=\linewidth]{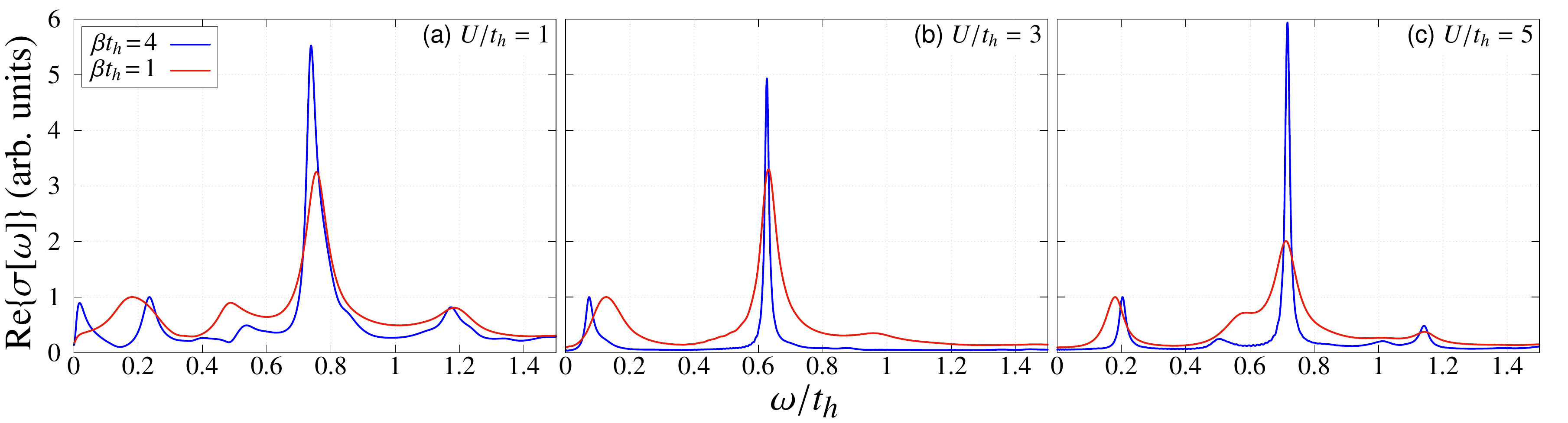}
  \caption{Same results as in Fig.~\ref{fig:OcEq}, but with all of the peaks  depicted.
  \label{fig:OcEqWhole}}
\end{figure*}  

\section{DERIVATION OF OPTICAL CONDUCTIVITY FOR EQUILIBRIUM STATE} \label{sec:appOpt}
Here, we illustrate the derivation of the optical conductivity for equilibrium systems, where external fields are set to zero: $A(t) = E_{\mu}(t) = 0$.
We omit the superscript $i$ from $t_{h}^{i}$, $U^{i}$, $g^{i}$, and $\Omega^{i}$ for the sake of simplicity. 
The current operator is defined as the functional derivative of the Hamiltonian with respect to the vector potential:
\begin{align}
\hat{j}(t) = -\frac{\delta \hat{H}_{S}(t) }{\delta A'(t)} = i\sum_{i, \sigma}(t_{h} e^{iA'(t)}\hat{c}_{i, \sigma}^{\dagger} \hat{c}_{i+1, \sigma} - \mathrm{H.c.}).
\end{align}
For sufficiently small $A'(t)$, we have $\hat{j}(t) \simeq \hat{j}^{p} - \hat{j}^{d} A'(t)$.
In the same way, the system Hamiltonian is expanded as 
\begin{align}
\hat{H}_{S} (t) \simeq{}&  -\sum_{i, \sigma} (t_h
\hat{c}_{i, \sigma}^{\dagger} \hat{c}_{i+1, \sigma} + \mathrm{H.c.})
+ \sum_{i} U \hat{n}_{i, \uparrow} \hat{n}_{i, \downarrow} \nonumber \\[1pt]
& + \sum_{i}g(\hat{n}_{i, \uparrow} + \hat{n}_{i, \downarrow})(\hat{b}_{i}^{\dagger} + \hat{b}_{i})
+  \sum_{i} \Omega \hat{b}_{i}^{\dagger} \hat{b}_{i}  \nonumber\\[1pt]
 &- \hat{j}^{p} A'(t) \nonumber \\[4pt]
= {}& \hat{H}^{0}_{S} - \hat{j}^{p} A'(t).
\end{align}
The total density operator for $\hat{H}_{\mathrm{tot}}$ is then expressed in the framework of  first-order perturbation theory for $ A'(t) $ as
\begin{align}
\hat{\rho}_{\mathrm{tot}}(t) = \hat{\rho}^{eq}_{\mathrm{tot}} + i \int_{-\infty}^{t} dt' \,G_{0}(t-t') [\hat{j}^{p}, \hat{\rho}^{eq}_{\mathrm{tot}}]A'(t').
\end{align}
Here, $G_{0}(t)\hat{O} = e^{-i\hat{H}_{\mathrm{tot}}^{0}t} \hat{O} e^{i\hat{H}_{\mathrm{tot}}^{0}t}$ is the time evolution operator without the perturbation defined as $\hat{H}_{\mathrm{tot}}^{0} \equiv \hat{H}_{S}^{0} + \hat{H}_{I} + \hat{H}_{B}$.   The expectation value of the current is then expressed in terms of the linear response function as \cite{mahan00, lenarcic14oc}
\begin{align}
\braket{\hat{j}(t)} &=  \tr \{\hat{j}(t) \hat{\rho}^{eq}_{\mathrm{tot}} (t)\} \nonumber \\
&=  \braket{\hat{j}^{p}}_\mathrm{eq} 
- \braket{\hat{j}^{d}}_\mathrm{eq} A'(t) 
 + \int_{-\infty}^{t} dt' R^{(1)}(t-t') A'(t), 
\label{eq:j}
\end{align}
where $R^{(1)}(t)= i \theta(t) \braket{[\hat{j}^{p}(t), \hat{j}^{p}(0)]}_\mathrm{eq}$ is the response function expressed in terms of the time-dependent current operator, defined as $\hat{j}^{p}(t) = e^{i\hat{H}_{\mathrm{tot}}^{0}t} \hat{j}^{p} e^{-i\hat{H}_{\mathrm{tot}}^{0}t}$, and $\braket{\hat{O}}_\mathrm{eq} = \tr\{\hat{O} \hat{\rho}^{eq}_{\mathrm{tot}}\}$ is the expectation value in the equilibrium state. Without external fields, the current does not flow and we have $\braket{j^{p}}_\mathrm{eq} = 0$.

Fourier expressions of $\braket{\hat{j}(t)}$, $A'(t)$, and $R^{(1)}(t)$ are expressed as $\braket{\hat{j}[\omega]}$, $A'[\omega]$, and $R^{(1)}[\omega]$, respectively.  Equation~\eqref{eq:j} is then expressed as 
\begin{align*}
\braket{\hat{j}[\omega]} = -\big(\!\braket{\hat{j}^{d}}  - R^{(1)}[\omega]\big)A'[\omega].
\end{align*}
Because the optical conductivity is defined as $\sigma[\omega] = \braket{\hat{j}[\omega]} / E'[\omega]$, and because the Fourier transform of the electric field $E'(t) = - \partial A'(t) / \partial t$ is expressed as $E'[\omega] = i \omega A'[\omega]$, we obtain Eq.~\eqref{eq:optEq}.

The optical conductivity is a response under external perturbation.  Thus, we exclude the equilibrium contribution $\braket{\hat{j}^{p}}_\mathrm{eq}$  
as in Eq.~\eqref{eq:optEq}, and the steady-state contribution $\braket{\hat{j}(t)}_{0}$ as in Eq.~\eqref{eq:opt}. 
For the direct evaluation of Eq.~\eqref{eq:opt}, we introduce an artificial damping factor $\epsilon$ in the Fourier transform of the vector potential, as $A'[\omega] = \int_{-\infty}^{\infty} dt\, e^{(i\omega + \epsilon) t} A'(t)$.
Equation~\eqref{eq:opt} is then modified and takes the following form:
\begin{align}
\sigma[\omega]  = \frac{1}{(i \omega + \epsilon) A'[\omega]}
\int_{0}^{\infty} dt \,e^{i \omega t}
\big[\braket{\hat{j}(t)} - \braket{\hat{j}(t)}_{0} \big].
\label{eq:optA}
\end{align}

When we set $A'(t) = A_{\mathrm{probe}}\theta(t-t_{\mathrm{on}})$, 
we obtain $A'[\omega] = -A_{\mathrm{probe}} e^{(i\omega + \epsilon) t_{\mathrm{on}}} / (i \omega + \epsilon)$.
By substituting this into Eq.~\eqref{eq:optA}, we obtain Eq.~\eqref{eq:optSimple} for $\epsilon \to 0$.

\section{ENTIRE PROFILES OF OPTICAL CONDUCTIVITY PRESENTED IN FIG.~\ref{fig:OcEq}} \label{sec:appPeaks}
Figure~\ref{fig:OcEqWhole} presents all of the peak profiles of optical conductivity calculated from Eq.~\eqref{eq:optEq}  for (a)  weak ($U/t_{h} = 1$), (b) intermediate ($U/t_{h} = 3$), and (c) strong ($U/t_{h} = 5$)  Coulomb repulsion in  the low-temperature case $\beta t_{h} = 4$ (blue curves) and the high-temperature case $\beta t_{h} = 1$ (red curves).  The bath parameters are given by $\gamma_{i}/t_{h} = 3$ and  $\eta_{i} = 0.05$ $(i=1, 2)$.  To conduct numerical calculations, we choose  $M_{LO} = 9$, $N_{\max} = 2$, and $K = 2$ for $\beta t_{h} = 4$, and $K = 1$ for $\beta t_{h} = 1$.  The low-frequency part, which is important for characterizing the conductivity, is presented in Fig.~\ref{fig:OcEq}.

\begin{figure}[!t]
\includegraphics[width=\linewidth]{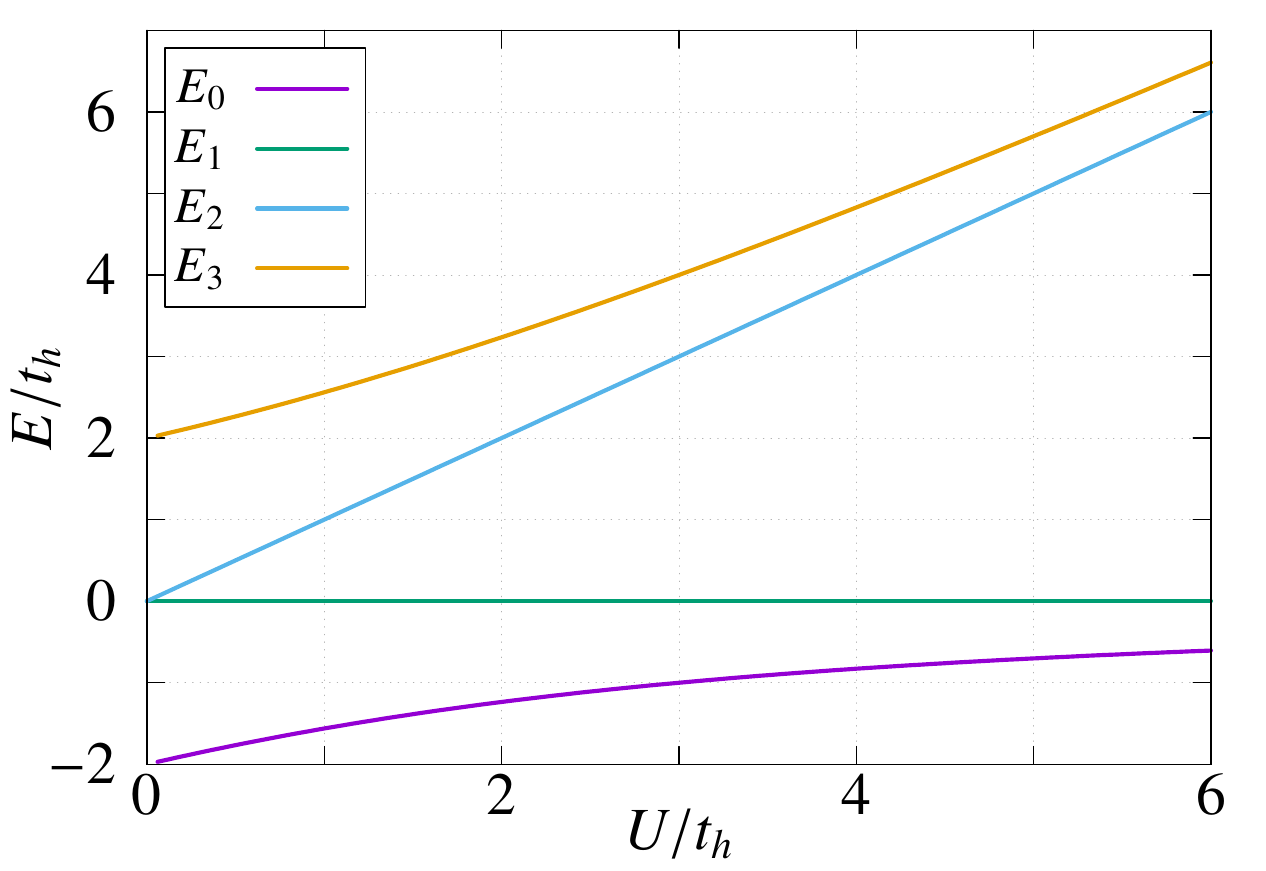}
\caption{Eigenenergies of half-filled two-site Hubbard model.
\label{fig:EneH}}
\end{figure}

We find that while the peak positions in the high-frequency regions do not change very much, regardless of the temperature, those in
the low-frequency region are temperature-dependent, as discussed in Sec.~\ref{subsec:eq}.  This is because these low-frequency peaks arise from the energy gap between the highest occupied and lowest unoccupied molecular orbitals (HOMO and LUMO). To illustrate this point, we consider the half-filled two-site Hubbard model whose eigenenergies are given by $E_{0} = U[1 - \sqrt{1 + (4t_h/U)^2}]/2$, $E_{1} = 0$, $E_{2} = U$,  and $E_{3} = U[1 + \sqrt{1 + (4 t_h /U)^2}]/2$. Here, $E_{1}$ corresponds to the triplet state, whereas the others correspond to the singlet states. The eigenenergies are depicted in Fig.~\ref{fig:EneH} as  functions of $U$.
Because the total spin number  $\hat{S}^{2} = [\sum_{i}(\hat{n}_{i, \uparrow} - \hat{n}_{i, \downarrow})/2 ]^{2}$ is conserved in our calculations, transitions between the singlet and triplet states do not occur.

In the low-temperature case ($\beta t_{h} = 4$), almost all populations are in the ground state. Thus, the major peak is due to the  $E_{0} \to E_{2}$ transition. In the large-$U$ region, the energy gap between these states is approximately $U$, while the effects from the LO phonons are minor: thus, in this region, the effective Coulomb repulsion can be estimated from the position of the maximum peak. Because the contribution of the LO phonons become larger when the Coulomb repulsion becomes smaller, the peak position and profile are not easy to predict, as explained by Fig.~\ref{fig:OcNeqE}.
The side peak around $0.5$ in the strong-repulsion case also results from the contribution of the LO phonons.
The side peak around $1$  originates from the oscillations of the LO phonons.

\section*{REFERENCES}

\bibliography{reference}
\end{document}